\documentstyle [12pt] {article}

\def\be{\begin{equation}}
\def\ee{\end{equation}}
\begin{document}
\begin{titlepage}
\title {\large \bf  Quantum Chains with $GL_q(2)$ Symmetry}
\author {Masoud Alimohammadi}
\maketitle
\begin {center}
{\it Institute for studies in Theoretical Physics and Mathematics
\\ P.O.Box 19395-5746 Tehran, Iran \\
 Department of Physics,Tehran University,\\
North Karegar,Tehran,Iran\\
email. alimohmd@irearn.bitnet }.
\end{center}
\vspace {10 mm}
\begin {abstract}
Usually quantum chains with quantum group symmetry are associated with
representations of quantized universal algebras $U_q(g) $ . Here we
propose
a method for constructing quantum chains with $C_q(G)$ global symmetry , where
$C_q(G)$ is the algebra of functions on the quantum group. In particular we
will construct a quantum chain with $GL_q(2)$ symmetry which interpolates
between two classical Ising chains.It is shown that the Hamiltonian
of this chain satisfies in the generalised braid group algebra.

\end{abstract}
\end{titlepage}
\pagebreak
\section {Introduction}

Almost all integrable models in two dimensional statistical models, quantum
field theories in 1+1 dimensions and qunatum chains $^{[1]}$ owe their
integrability to
some quantum group symmetry. For example in lattice models , if one assigns the
local Boltzman weights of a vertex or an IRF model to be the elements of the R
matrix corresponding to a quantum group, then the model will be integrable due
to the existence of a one parameter family of commuting transfer matrices. In a
sence one can say that local quantum group symmetry ensures integrability.
Although global quantum group symmetry does not mean integrability ,
construction of
models with such symmetries may be interesting and important as a first step
toward understanding the mechanism of integrablity.

Recently new types of 2 and 3 states quantum chains were constructed and shown
to possess $ U_q(sl(2)) $ symmetry $^{[2-4]}$. The strategy followed in [4]
was to define the Hamiltonian as
\be H = \sum _{j=1}^L id\otimes ....id\otimes  H_j \otimes id ....\otimes id
\ee
where $ H_j$  acts on sites $j$ and $ j+1$ as
\be H_j = ( \pi_j \otimes \pi_{j+1} ) [ Q_j ( \Delta ( C ) ) ]. \ee
Here $ j $ denotes the site of the lattice ,
$ C $ is the  quadratic Casimir of $ U_q(sl(2)) $ , $ \Delta$ is the
coproduct , $ \pi_j $ is a typical type
$ b $ representation $^{[5]}$ of $ U_q(sl(2)) $  assigned to site $ j $ , and
finally $ Q_j $ is a polynomial function of
degree $ d\leq m $  where the integer $ m $  is characterized by the value of
$ q $ ($  q^m = 1 $) .
This Hamiltonian is by construction $ U_q(sl(2)) $ invariant . The
invariance is due to
the centrality of the Casimir. For the particular form of the Hamiltonian of
the 2-states and 3-states quantum chains see Ref.
[4].

As is well known any quantum group is characterized by two algebras $^{[6,7]}$.
The first
being the deformation of the universal enveloping algebra which is denoted by
$ U_q(g)$ and the second one which is the deformation of the algebra of
functions on the group which is denoted by $ C_q(G)$.
So far everything which has been done concerning the construction of physical
models with quantum group symmetry have been based on representation theory of
$ U_q(g)$. However in the quantum case the second algebra , $ C_q(G)$ ,
has also
a representation theory which is completely different
from that of $ U_q(g)$ $^{[8,9]}$ .

The novelty of the representations of $ C_q(G)$ is best understood when one
considers the classical $ (q \longrightarrow 1) $ limit of $ C_q(G) $ . In this
limit,
representations of $ U_q(g) $ approach those of the classical Lie algebra  $ g
$ while those of $ C_q(G)$ collapse to trivial one dimensional
representations, since $ C_q(G)$ will become a commutative algebra.
Therefore there is no parallelism between the representation theories in the
deformed and the undeformed case. Naively one expects that paying attention to
physical models which are $ C_q(G)$ invariant may open up a new road in the
study of integrable models. At the present stage this is only a hope and real
justification for it will exist if one can somehow gauge a global
symmetry of this kind in a particular physical model .

In this letter we construct a quantum chain which has  a global $ C_q(GL(2)) $
symmetry , hereafter called $ GL_q(2)$ symmetry for simplicity.

\section {The quantum group $ GL_q(2) $ and it's cyclic representations }

The quantum
group $GL_q (2)$ is defined by the generators  $ 1,a,b,c $ and $d $, collected
in the form of a quantum matrix $^{[7]}$
$$ T = \left( \begin{array}{ll} a&b
\\ c&d
\end{array} \right) $$
and relations
   $$ab=qba \hskip 2cm  ac=qca  $$
   $$bd=qdb \hskip 2cm  cd=qdc  $$
\be bc=cb \ee
   $$ad-da=(q-q^{-1})bc$$
   The coproduct which is used in tensor multiplication of representations is
   defined by :

\be \Delta  \left( \begin{array}{ll} a&b \\ c&d\end{array} \right) =
\left( \begin{array}{ll} a&b \\ c&d\end{array} \right) \otimes
\left( \begin{array}{ll} a&b \\ c&d\end{array} \right) \ee \\
The quantum determinant $D_{q}=ad-qbc$ is central and group-like, That is:
\be \Delta D_q=D_q \otimes D_q \ee
If q is a root of unity ( $ q^p =1 ) $ , in addition to the
determinant, all the
elements $ a^p, b^p , c^p $ and $ d^p $ are central. In this case the algebra
has a $ p $ dimensional cyclic representation which is constructed as follows
$^{[8]}$ : One first defines a state $ \vert 0 > $  which is a common
eigenvector
of $ b$  and $ c $  with eigenvalues $ \mu $ and $ \nu $  respectively:
$$ b \vert 0 > = \mu \vert 0 > \hskip 2cm   c \vert 0 > = \nu \vert 0 >$$
and then builds up the representation space $ V $  as the linear span of
vectors $ \{ \vert n > \equiv d^n \vert 0 > \hskip 1cm 0 \leq n \leq p-1  \}
$ . It is then easy to show that $ V $ is invariant under the action of $
GL_q(2)$.

$$d \vert n> = \vert n+1> ,  \hskip 1cm   d \vert p-1>= \eta \vert 0> $$
$$b \vert n> =\mu q^n \vert n> , \hskip 1cm  c \vert n> =\nu q^n \vert n> $$
\be a \vert n>=(\xi + q^{2n-1} \mu \nu) \vert n-1> \ee
$$a \vert 0>={1 \over \eta} (\xi + \mu \nu q^{-1}) \vert p-1>$$
here $\eta$ is the central value of $d^p$ and $\xi$ is the value of the
q-determinant $D_q$.

It can be easily checked that the parameters $ \eta , \xi ,\mu $ and $ \nu $
are all independent and hence each representation is characterized by the
values of these parameters and is denoted by $ \pi (\eta , \xi ,\mu , \nu )$.\\

 \section { Quantum chains with $GL_q(2)$ symmetry }

We construct the quantum chains with $ GL_q(2) $ symmetry as follows:
To each site $ 1\leq j \leq L $ , we assign a representation
$ \pi_j = \pi (\eta_j , \xi_j ,\mu_j ,\nu _j)$ , the Hilbert space is the
tensor
product $ \otimes^{L}_{j=1} V_j $ , where $ V_j $ is the $ p $- dimensional
representation space of $ \pi_j $ .

At first glance it seems that the analogue of the
construction
of Ref.[4] in the case of $ C_q(GL(2))$ is to replace the Casimir C
in eq.(2) by the
quantum determinant $D_q$. However this procedure leads to a trivial
Hamiltonian due to the group like property of $D_q$ (eq.(5)), which makes
$\pi_j \otimes \pi_{j+1} (Q_j ( \Delta (D)))$ proportional to the identity.
However there is one interesting possibility and it is to define $H_j$ as:
\be H_j=\pi _j \otimes \pi _{j+1} (Q_j(\Delta (a^p), \Delta (b^p), \Delta
(c^p),\Delta (d^p))) \ee
Here
the crucial point is that  although in an irreducible representation
$a^p,b^p,c^p$ and $d^p$ are proportional
to the identity, in a tensor product of representations they are not so due to
the  mixing of the generators in their coproducts (eq.(4)).
The Hamiltonian constructed in this way is $ GL_q(2) $ invariant by
construction.\\  \\

{ \bf Two state quantum chains }\\

Now we restrict ourselves to the case $p= 2 \ \ (q = -1)$. From eq.(6)
we obtain the 2-dimensional cyclic representation of $GL_q(2)$, which in the
explicit matrix notation takes the form:
$$ \pi ( a ) = \left( \begin{array}{ll} 0&{\gamma / \eta }
\\ {\gamma}&0
\end{array} \right)\hskip 2cm
\pi (b)  = \left( \begin{array}{ll} \mu&0
\\ 0&{-\mu}
\end{array} \right)$$\\
\be \pi (c) = \left( \begin{array}{ll} {\nu}&0
\\ 0&-\nu
\end{array} \right)\hskip 2cm
\pi ( d ) = \left( \begin{array}{ll} 0&1
\\ \eta & 0
\end{array} \right) \ee
where $\gamma = \xi - \mu \nu $.
This represent a continuous 4-parameter family of two dimensional
representations for $GL_q(2)$ (for $q=-1$).
If $t$ stands for $ a,b,c$ or $d$ , then
a straightforward calculation shows that:
\be (\pi_j \otimes \pi_{j+1})\Delta t^2=
l_t1\otimes 1+n_t \sigma_x \otimes \sigma_x
+p_t \sigma_y \otimes \sigma_y -iq_t \sigma_x \otimes \sigma_y
-ir_t \sigma_y \otimes \sigma_x \ee
where
$$n_t={1\over 4}(1-\eta_j -\eta_{j+1}+\eta_j\eta_{j+1})m_t$$
$$p_t=-{1\over 4}(1+\eta_j +\eta_{j+1}+\eta_j\eta_{j+1})m_t$$
\be q_t={1\over 4}(-1+\eta_j -\eta_{j+1}+\eta_j\eta_{j+1})m_t \ee
$$r_t={1\over 4}(-1-\eta_j +\eta_{j+1}+\eta_j\eta_{j+1})m_t$$
and
$$m_a={{2\gamma_j \gamma_{j+1}\mu_j\nu_{j+1}}\over {\eta_j\eta_{j+1}}}
\hskip1cm
, \hskip1cm m_b=-2{{\mu_j\mu_{j+1}\gamma_j} \over {\eta_j}}$$
\be m_c=-2{{\nu_j\nu_{j+1}\gamma_{j+1}} \over {\eta_{j+1}}} \hskip1cm ,
\hskip1cm m_d=2\mu_{j+1}\nu_j \ee
The explicit expressions of $l_t$'s are not necessary in this stage.
As the simplest choice for the polynomial (in eq.(7)) we set :
\be Q_0= \alpha_aa^2+\alpha_b b^2+\alpha_c c^2+\alpha_d d^2 \ee
where $\alpha_t$'s are arbitrary constants.
Combination of eqs.(9) and (12) leads to the following Hamiltonain:
\be H_j=A\sigma_x^j\sigma_x^{j+1}+ B\sigma_y^j\sigma_y^{j+1}+
-iC\sigma_x^j\sigma_y^{j+1}-i D\sigma_y^j\sigma_x^{j+1} \ee
where
$$A=\beta_j(1-\eta_j -\eta_{j+1}+\eta_j\eta_{j+1})$$
$$B=-\beta_j(1+\eta_j +\eta_{j+1}+\eta_j\eta_{j+1})$$
\be C=\beta_j(-1+\eta_j -\eta_{j+1}+\eta_j\eta_{j+1}) \ee
$$D=\beta_j(-1-\eta_j +\eta_{j+1}+\eta_j\eta_{j+1})$$
and $\beta_j=\alpha_am_a+\alpha_bm_b+\alpha_cm_c+\alpha_dm_d$.
If the factor $\beta_j$ is site independent , then  modulo a constant overall
factor the Hamiltonian becomes:
$$ H_0=\sum_j \{ (1-\eta_j -\eta_{j+1}+\eta_j\eta_{j+1})\sigma_x^j
\sigma_x^{j+1}
- (1+\eta_j +\eta_{j+1}+\eta_j\eta_{j+1})\sigma_y^j \sigma_y^{j+1}$$
\be -i (-1+\eta_j -\eta_{j+1}+\eta_j\eta_{j+1})\sigma_x^j \sigma_y^{j+1}
-i (-1-\eta_j +\eta_{j+1}+\eta_j\eta_{j+1})\sigma_y^j \sigma_x^{j+1} \} \ee
 Now the condition of Hermiticity of the Hamiltonian restricts the parameters
$\eta_j$ and $\eta_{j+1}$ to the following form:
$$\eta_j=\alpha +i\sqrt{1-\alpha^2}$$
\be \eta_{j+1} = \eta_j^*, \ee
where $\alpha$ is a real parameter.Under this condition the Hamiltonian takes
the following simple form:
\be H_0=\sum_j \{(1-\alpha) \sigma_x^j \sigma^{j+1}_x-
(1+\alpha) \sigma_y^j \sigma^{j+1}_y+
\sqrt{1-\alpha^2}( \sigma_x^j \sigma^{j+1}_y-
 \sigma_y^j \sigma^{j+1}_x)\}. \ee
This is the desired Hamiltonian with $GL_q(2)$ symmetry.

Imposing the condition of site independence on $\beta_j$ in eq.(14)
, the $m_t$'s are restricted to be site independent , as $\alpha_t$'s are
arbitrary
constants. Solving these conditions and using eq.(16) , results the following
relation between the parameters of the different representations of the sites:
$$\gamma_j=\gamma_{j+2} \hskip1cm,\hskip1cm \nu_j=\nu_{j+2} \hskip1cm,
\hskip1cm \mu_j=\mu_{j+2} $$
\be {\gamma_j \over \eta_j} = {\gamma_{j+1}\over \eta_{j+1}} \hskip15mm ,
\hskip15mm {\mu_j \over \nu_j} = {\mu_{j+1} \over \nu_{j+1}} .\ee
So the whole representations of the sites specify only by four complex
parameters
$\nu_1,\mu_1,\nu_2,\gamma_1$ and one real parameter $\alpha$.

There are several observations on the above Hamiltonian (eq.(17)):\\
{\bf 1)} Instead of the original continuous parameter $q$ , the Hamiltonian
depends
on the continuous parameter $\alpha$, which comes from the representation.\\
{\bf 2)} In the two limits $\alpha=1$ and $\alpha=-1$, this Hamiltonian
degenerates into
a exactly solvable chain , that is $ \sum ^L_{j=1} \sigma ^j _{\hat n} \sigma
^{j+1} _{\hat n}$ .
So our $GL_q(2)$ invariant Hamiltonian interpolates between two {\bf xx }and
{\bf yy} classical Ising chains, when $\alpha$ is changed continuously from
$-1$
to
$1$. \\ {\bf 3)} It is crucial to note that the Hamiltonian (17) is not
equivalent
to an ${\hat n} {\hat n}$ chain where ${\hat n}$ is a new unit vector in the
$x-y$ plane. That
is the transformations $\sigma_x \longrightarrow a \sigma_x + b \sigma_y $ and
$\sigma_y \longrightarrow c \sigma_x + d \sigma_y $ can not diagonalize the
Hamiltonian. \\
{\bf 4)} If one defines $U_i=2+H_i$ , then there exists the following
interesting
relation between $U_i$'s:
$$U_i^2=4U_i$$
\be (U_iU_{i\pm 1}U_i - U_{i\pm 1}U_iU_{i\pm 1})(U_i-U_{i\pm
1})=64(1-\alpha^2).\ee
The above equation is the generalised braid group algebra $^{[2]}$.\\
{\bf 5)} The simplest situation that $H_0$ can be solved exactly is the case of
2-site lattice. In this case there are four states with eigenvalues $2,2,-2$
and $ -2$.
It can be shown that the degenarate orthonormal states ($\vert \ \  2 >_\pm $
and
$\vert -2>_\pm$) , are two-dimensional representations of $\Delta (t)$'s.\\

{\bf Higher state quantum chains} \\

Choosing p=3, one may expect to obtain a 3-state $GL_q(2)$ invariant quantum
chain. However, for $q^3=1$ it is seen by computation that:
$$\Delta a^3=a^3 \otimes a^3+b^3 \otimes c^3$$
$$\Delta b^3=a^3 \otimes b^3+b^3 \otimes d^3$$
$$\Delta c^3=c^3 \otimes b^3+d^3 \otimes d^3$$
$$\Delta d^3=c^3 \otimes a^3+d^3 \otimes c^3$$
which means that the Hamiltonian (7) is proportional to the
identity.This phenomena may occur for all odd integers $p$.\\

{\bf Acknowledgement } I would like to thank V.Karimipour for useful
discussions. \\

{\large \bf References}     \\
\begin{enumerate}

\item  For a review see  H. Saluer and J. B. Zuber : " Integrable Lattice
Models
and Quantum Groups " Proceedings of the Trieste Spring School (1990), World
Sicentific, Singapore
\item H. Hinrichsen and V. Rittenberg : Phys. Lett. {\bf B } 275 (1992) 350
\item H. Hinrichsen and V. Rittenberg : Phys. Lett. {\bf B } 304 (1993) 115
\item D. Arnaudon and V. Rittenberg : Phys. Lett. {\bf B } 306 (1993) 86
\item P. Roche and D. Arnaudon: Lett. Math. Phys. {\bf 17} (1989) 295
\item V. G. Drinfeld " Quantum Groups ", Proc. Internat. Congr. Math. (
Berkeley) ,vol. 1, Academic Press ,New York, 1986
\item  N. Yu. Reshetikhin , L.A. Takhtajan , and L.D. Faddeev ; Lenningrad
Math. Journal  Vol. 1 (1990) 193
\item  M. L. Ge, X. F. Liu , and C. P. Sun : J. Math. Phys. 38 (7) 1992
\item  V. Karimipour; Lett. Math. Phys. {\bf 28} 1993,207  \\
 V. Karimipour; J. Phys. A: Math. Gen. {\bf 26} (1993) 6277

\end {enumerate}

\end {document}